\documentclass[aps,prl,twocolumn,showpacs,groupedaddress]{revtex4}

\usepackage{times}
\usepackage{graphicx}
\usepackage{amsmath}
\newcommand{\Mdata}{M_{\text{data}}}
\newcommand{\Mrot}{M_{\text{rot}}}
\newcommand{\Mpix}{M_{\text{pix}}}
\newcommand{\qmax}{q_{\text{max}}}
\newcommand{\qmin}{q_{\text{min}}}

\makeatletter
\newif\if@restonecol
\makeatother

\begin{document}

\title{Cryptotomography: reconstructing 3D Fourier intensities from randomly oriented single-shot diffraction patterns} 

\author{N. D. Loh$^1$}
\author{M. J. Bogan$^2$}
\author{V. Elser$^1$}
\author{A. Barty$^3$}
\author{S. Boutet$^2$}
\author{S. Bajt$^4$}
\author{J. Hajdu$^5$}
\author{T. Ekeberg$^5$}
\author{F. R. N. C. Maia$^5$}
\author{J. Schulz$^3$}
\author{M. M. Seibert$^5$}
\author{B. Iwan$^5$}
\author{N. Timneanu$^5$}
\author{S. Marchesini$^6$}
\author{I. Schlichting$^{7,\,8}$}
\author{R. L. Shoeman$^{7,\,8}$}
\author{L. Lomb$^{7,\,8}$}
\author{M. Frank$^9$}
\author{M. Liang$^3$}
\author{H. N. Chapman$^{3,\,{10}}$}
\affiliation{$^1$Laboratory of Atomic and Solid State Physics Cornell University, Ithaca, NY 14853-2501, USA\\
$^2$SLAC National Accelerator Laboratory, 2575 Sand Hill Road, Menlo Park, California, 94025, USA\\
$^3$Center for Free-Electron Laser Science, DESY, Notkestrasse 85, Hamburg 22607, Germany\\
$^4$Photon Science, DESY, Notkestrasse 85, Hamburg 22607, Germany\\
$^5$Laboratory of Molecular Biophysics, Department of Cell and Molecular Biology, Uppsala University, Husargatan 3, Box 596, SE-75124 Uppsala, Sweden\\
$^6$Lawrence Berkeley National Laboratory, 1 Cyclotron Road, Berkeley CA 94720, USA\\
$^7$Max Planck Institute for Medical Research, Jahnstr. 29, 69120 Heidelberg, Germany\\
$^8$Max Planck Advanced Study Group, Center for Free-Electron Laser Science, DESY, Notkestrasse 85, Hamburg 22607, Germany\\
$^9$Lawrence Livermore National Laboratory, 7000 East Avenue, Livermore, CA 94550, USA\\
$^{10}$University of Hamburg, Luruper Chaussee 149, Hamburg 22761, Germany
}

\date{\today}

\begin{abstract}
We reconstructed the 3D Fourier intensity distribution of mono-disperse prolate nano-particles using single-shot 2D coherent diffraction patterns collected at DESY's FLASH facility when a bright, coherent, ultrafast X-ray pulse intercepted individual particles of random, unmeasured orientations. This first experimental demonstration of cryptotomography extended the Expansion-Maximization-Compression (EMC) framework to accommodate unmeasured fluctuations in photon fluence and loss of data due to saturation or background scatter. This work is an important step towards realizing single-shot diffraction imaging of single biomolecules.
\end{abstract}


\maketitle


Single-shot coherent diffraction imaging (CDI) is a developing technique aimed at determining the structure of very small biomolecules, such as proteins and viruses, not easily imaged through more established methods. In one scheme of single-shot CDI a serial stream of particles is injected into a pulse train of highly coherent and energetic X-ray free electron laser (FEL) radiation \cite{Bogan2010}. Photons from a single FEL pulse are diffracted when they encounter a particle. Although the FEL pulse destroys this particle, its structure is recorded in the diffraction data if the pulse is sufficiently short \cite{Neutze, Chapman2006}. 

As it is difficult to manipulate or determine the orientations of very small particles, they are currently injected into the FEL radiation at {\it random}, unmeasured orientations. Nevertheless, sufficiently many 2D diffraction patterns from an ensemble of identical albeit randomly oriented particles can in principle overdetermine the particle's band-limited 3D Fourier intensities. Earlier numerical simulations also show that the particle's 3D intensities can be recovered even if the random 2D diffraction patterns are remarkably {\it noisy} \cite{Elser, Loh, Ourmazd}. 

\begin{figure}[t!]
\centering
\includegraphics[width=2.1in]{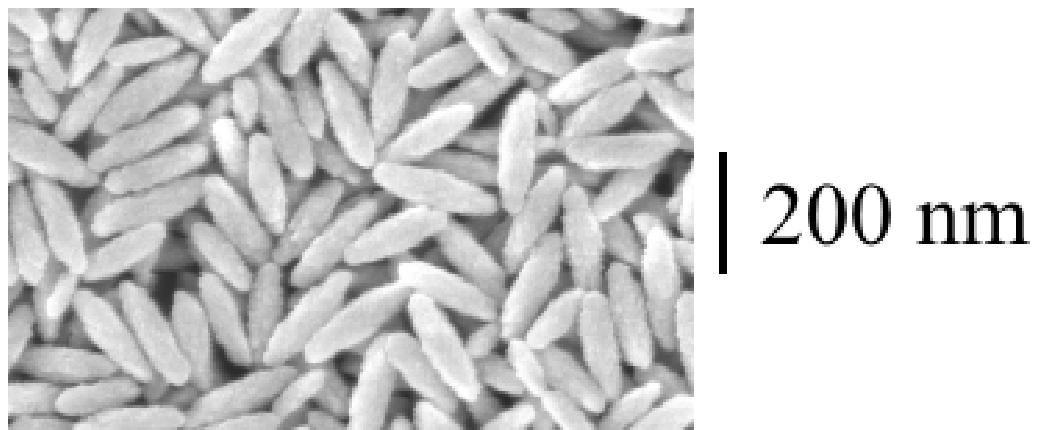}
\includegraphics[width=1.6in]{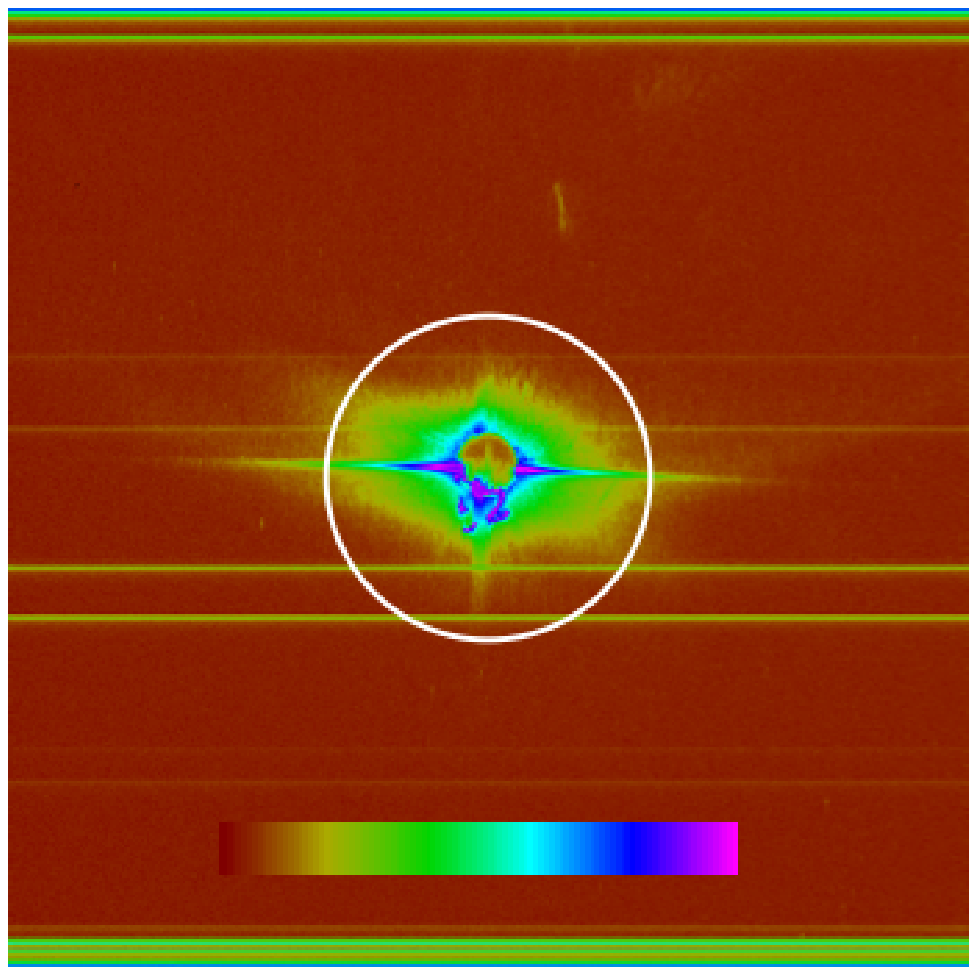}
\includegraphics[width=1.6in]{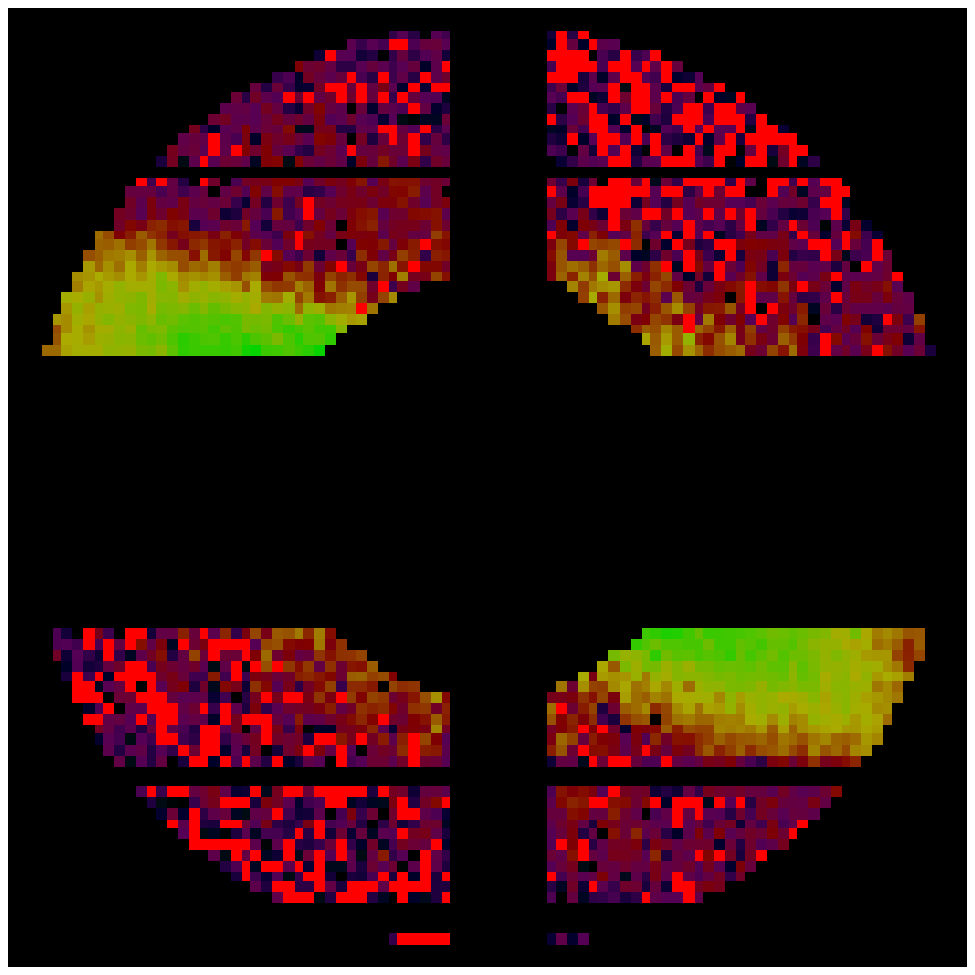}
\caption{(Color online) Top panel shows an SEM image of the prolate iron oxide nano-particles (oblate 3D Fourier intensities) used in our experiment. Lower panels show a typical noisy and random diffraction pattern from such a nano-particle before ($512\times512$ array, left) and after processing ($91\times91$ array, right): truncated high spatial frequencies (discarded photons counts outside left white circle); binned photon counts; subtracted background; excised non-signal pixels (redacted). The logarithm of photon counts are colored according to the inset color bar (max. at pink); negative counts in red. The mean dynamic range of photon counts in these post-processed data spans 2 orders of magnitude, subject to fluctuations in fluence. \label{fig:data}}
\end{figure}

This Letter demonstrates the experimental feasibility of single-shot CDI with noisy 2D diffraction patterns from randomly-oriented identical particles, which we coin {\it cryptotomography}. In this first exercise in cryptotomography, we reconstruct the 3D intensities of a relatively large and simple iron oxide nano-particle. These nearly mono-disperse particles are approximately solids of revolution with principal radii 25 nm and 100 nm (SEM measurements shown in Fig. \ref{fig:data}). 

Our primary objective is not to study these simple nano-particles in any greater detail than what is already available via SEM, but to show that cryptotomography, despite its random and noisy data, is experimentally viable through the union of experimental and theoretical innovations. We do so in the style of previous papers in diffraction imaging, which demonstrated novel techniques with simple test subjects \cite{Chapman2006, GoldPyramid}.

We performed our experiment at DESY's FLASH facility with each FEL pulse train comprising 100 pulses (7 nm radiation; $30 \pm 10 \mu$m beam focus, extrapolated from \cite{beamSize}) separated by 10 $\mu$s, repeated at 5 pulse trains per second. Each detector exposure was 1 second long. Additional details on experimental parameters and data collection were similar to \cite{Bogan2010} and \cite{Bogan2009, multilayerMirror} respectively, except we used a nebulizer instead to aerosolize the nano-particles, which were then directed into the FEL radiation using an aerodynamic lens stack \cite{lensStack}. 
 
Our experiment imposed two considerable challenges. First, the FEL pulse fluence fluctuated due to electron bunch dynamics in the lasing process \cite{Saldin}. Second, there was considerable diffuse background scattering (Fig. \ref{fig:data}) from a silicon aperture used to shield downstream instruments from beamline scatter. All 5$\times$100 pulses in a single detector exposure contributed to this scatter whereas only 1 incident pulse creates a useful diffraction pattern if it illuminates a particle. This diffuse background changed gradually as the FEL pulses eroded the edges of the aperture, exacerbating this issue. 

To overcome the second challenge, we assumed that the background scatter added incoherently to the diffraction pattern from a pulse illuminating a particle. From 2000 diffraction patterns (data) we identified those that contained coherent scattering (hits), by checking for intensity lobes expected of single prolate nano-particles (Fig. \ref{fig:data}) while excluding anomalous data with scattering from injected debris or those with diffraction patterns from more than one particle. Only non-hits without anomalies were considered background data.

Since the character of the background changed slowly over many data-acquisition cycles, for each hit we compared the results after separately subtracting five background data acquired nearest in time. From these five subtractions we selected the one that gave roughly equal numbers of positive and negative photon counts at higher spatial frequencies, where the particle's signal is presumably negligible \footnote{Some potential hits still had strong pure background, due to the mismatched total fluence of potential hits and compared background data, and had to be discarded.}. 

Row defects, scattering from the aperture in the multi-layer planar mirror \cite{multilayerMirror} and missing photon counts at the lowest spatial frequencies were always limited to certain pixels of the X-ray sensitive detector. These pixels were excised from all data --- their measurements did not constrain intensity reconstruction. 2880 ($\Mpix$) usable pixels remained, identical in all hits (non-redacted pixels in Fig. \ref{fig:data}, lower right).


Each of the resultant 54 ($\Mdata$) background-subtracted hits, despite missing information about their orientation ($\Omega$) and fluence ($\phi$), are noisy Ewald sphere sections of the particle's true 3D intensities. Our goal: to recover the set of 3D intensities and fluences most statistically compatible with these hits.  



Our algorithm for recovering the particle's 3D intensities \cite{Elser, Loh} is based on expectation maximization \cite{EM}, where we iteratively apply a simple rule to increase the compatibility of any model intensities ($W$, even random ones) with all hits. Consider the simplest case where we are given only one hit. We use a statistical test to determine which Ewald sphere sections in $W$, here on known as {\it tomograms}, are most compatible with this hit, then replace those tomograms with said hit, weighted by probabilities taken from the statistical test. This prescription on $W$, which is iterated to a fixed point, determines the most likely orientations of this hit with respect to an increasingly compatible $W$. We can generalize this prescription to many hits by updating the tomograms in consensus with all hits. This Letter extends our algorithm to also determine the fluence distribution of the hits.

The model 3D intensity $W$ has various representations. It can be compactly written as $W(\mathbf{q})$: the time-integrated scattered intensity at spatial frequency $\mathbf{q}$ when the particle is in some reference orientation. We represented $W(\mathbf{q})$ in our reconstructions as a cubic array of floating-point numbers, indexed by equally spaced samples of ${\mathbf{q}}$. Detector pixels, labeled by index $i$, measure $M_\mathrm{pix}$ point samples $W(\mathbf{q}_i)$. 

If we gave the particle some arbitrary orientation $\Omega$ from the reference position, the intensity recorded at detector pixel $i$ is $W(\mathbf{R}_\Omega\cdot\mathbf{q}_i)$, where $\mathbf{R}_\Omega$ is the orthogonal matrix of this rotation. We approximated the continuous $\Omega$ with a discrete sampling of $\Mrot$ points labeled by the index $j$ (sampling discussed in \cite{Loh}). As a shorthand, we define the intensity at detector pixel $i$, after dividing away the incident pulse fluence, from diffraction off a particle at approximate orientation $j$ as tomogram samples $W_{ij} = W(\mathbf{R}_j\cdot\mathbf{q}_i)$.

The statistical test central to our expectation-maximization algorithm assess the likelihood that measurements on the $i$th detector pixel of the $k$th hit ($K_{ik}$), with photon fluence $\phi_k$, corresponds to tomogram samples of our intensity model $W_{ij}$. Intensity fluctuations at each detector pixel of a particular hit due to background-subtraction are assumed to be mutually independent. Assuming the noise from background-subtraction dominates over Poisson statistics, the likelihood that the $k$th hit comes from the $j$th tomogram of $W$ is
\begin{equation}
R_{jk}(W,\Omega, \phi) \propto \exp{ \left( - \, \frac{\sum_i^{\Mpix} (K_{ik}/\phi_k - W_{ij})^2}{2 \; \sigma^2} \right)} \;  . \label{eqn:Gaussian}
\end{equation}
The global noise parameter $\sigma$ in eq. \eqref{eqn:Gaussian} is the only unconstrained parameter in our algorithm. We seek an ideal $\sigma$ that quantifies the true noise in the diffraction data. 


Determining the most likely model parameters $(W, \Omega, \phi)$ given all hits is unattainable in a single step, hence we adopt an iterative procedure that we call Expansion-Maximization-Compression (EMC) to implement the model updates \cite{Loh} . 

In the EMC prescription we first expand (E-step) $W({\bf q})$ into the set of $\Mrot$ tomograms $W_{ij}$. Although this expansion allows efficient comparisons between hits and model tomograms, it creates redundancy since each intensity sample of the 3D model $W({\bf q})$ is represented in multiple tomograms $W_{ij}$. Such redundancy ameliorates the effects of pixel excisions when populating $W({\bf q})$ with data measurements.

After expanding our model $W$ into tomograms, we maximize (M-step) the data likelihood of each model tomogram $W_{ij} \to W^\prime_{ij}$ independently. Specifically, we determine the new model $(W_{ij}^\prime, \Omega^\prime_k, \phi^\prime_k)$ which maximizes Eq. \eqref{eqn:EMC}, conditional on probabilities of the current model $(W_{ij}, \Omega_k, \phi_k)$:  
\begin{equation}
\underset{W^\prime, \Omega^\prime, \phi^\prime}{\text{argmax}} \sum_k^{\Mdata} \sum_j^{\Mrot} R_{jk}(W,\Omega, \phi)\, w_j \; \log \left(  R_{jk}(W^\prime,\Omega^\prime,\phi^\prime)  \right) \label{eqn:EMC}\; .
\end{equation}
The numbers $w_j$ are weights applied to our rotation group samples that approximate a uniform prior distribution \cite{Loh}. Despite this uniform prior distribution over the rotation group, we can still detect orientational biases by evaluating Eq. \eqref{eqn:Gaussian} for all hits over the converged $W$ reconstructions.

The requirement that $W_{ij}^\prime$ from different orientations are consistent with a single intensity model $W^\prime({\bf q})$ is enforced in the final compression (C-step) by averaging interpolated intensity samples in all tomograms (details in \cite{Loh}), which represent a particular intensity sample in $W^\prime({\bf q})$. We also impose Friedel symmetry on $W^\prime({\bf q})$ since we are operating in the limit of weak elastic scattering. This compressed and symmetrized model $W^\prime({\bf q})$ is now ready for another round of EMC.

We exploited a side-effect of EMC's redundant intensity representation to find the ideal noise parameter $\sigma_{\text{min}}$. If $\sigma$ is too small, even though the data likelihood of each updated tomogram is provisionally increased, such tomograms are mutually incompatible, thus diminishing the data likelihood of the compressed updated intensities. We determined $\sigma_{\text{min}}$ knowingly: if $\sigma < \sigma_{\text{min}}$, the log-likelihood of reconstructions, Eq. \eqref{eqn:EMC}, decline and reconstructions vary dramatically.

EMC updates of the fluence and intensity model have a regrettable degeneracy: if scaling $W^\prime({\bf q})$ by a multiplicative constant increases the model's log-likelihood in \eqref{eqn:EMC} so will a commensurate scaling in $\phi^\prime$. As a consequence, simultaneous EMC updates for $\phi_k$ and $W_{ij}$ cannot be decoupled easily. However, if we updated only $\phi_k$ or $W_{ij}$, while keeping the other fixed, the net log-likelihood still increases:

\begin{eqnarray}
W_{ij}^\prime &=& \frac{\sum_k R_{jk} w_j \; K_{ik} / \phi_k}{ \sum_k R_{jk} w_j} \label{eqn:WUpdate} \, , \\
\phi^\prime_k  &=& \frac{\sum_j R_{jk} w_j \; \sum_i K_{ik}^2}{\sum_j R_{jk} w_j \; \sum_i K_{ik} W_{ij}} 
\label{eqn:phiUpdate} \, .
\end{eqnarray}
The likelihood $R_{jk}$ in the last two equations is evaluated at the current model parameters $(W, \Omega, \phi)$. We imposed $\sum_j R_{jk} w_j= 1$ during each EMC iteration to assert that every hit (index $k$) must be found at some orientation (index $j$). 


\begin{figure}[t!]
\centering
\includegraphics[width=1.1in]{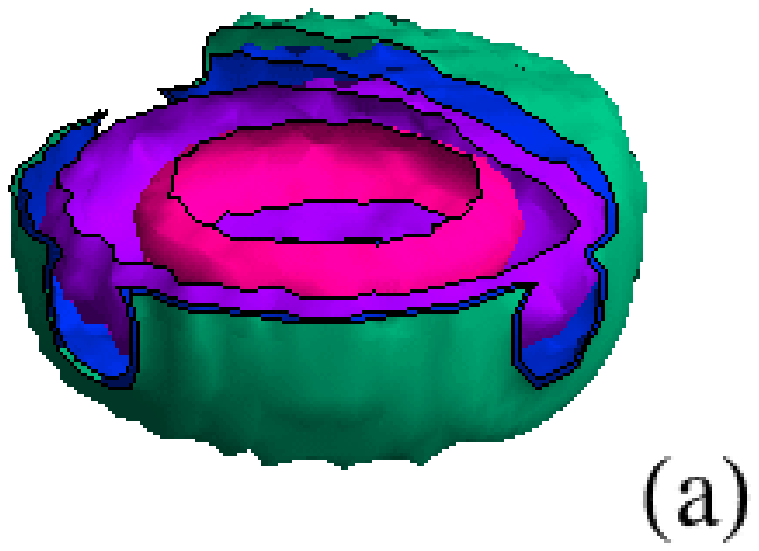}
\includegraphics[width=1.2in]{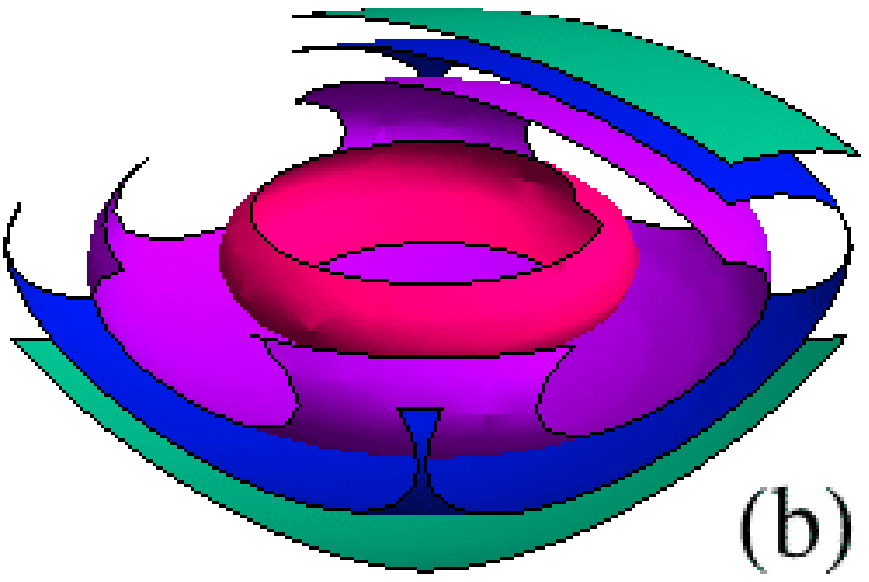}
\includegraphics[width=3.in]{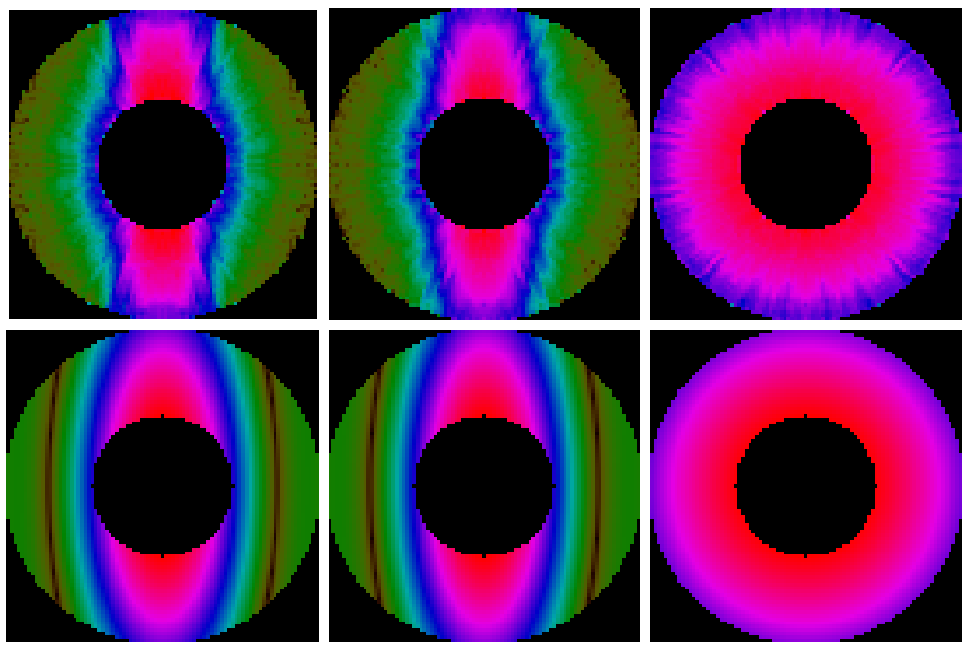}
\caption{(Color online) Comparing reconstructed intensities to those of an ideal ellipsoidal particle $I_{\text{ellip}}$. (a) Cutaway view of choice 3D iso-intensity surfaces of a reconstruction which show an oblate intensity distribution and (b) those of $I_{\text{ellip}}$ with the best R-factor fit to this reconstruction; (middle row) mutually perpendicular cross-sections of this reconstruction; (bottom row) same cross-sections of $I_{\text{ellip}}$ in (b). Logarithm of intensities are shown as hues (color bar in Fig. \ref{fig:data}). Intensities in reconstructions span 3 orders of magnitude. \label{fig:recon}}
\end{figure}

\begin{figure}[t!]
\centering
\includegraphics[width=1.9in]{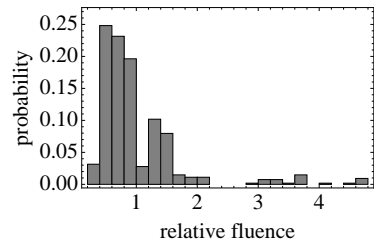}
\caption{Probability distribution of reconstructed relative fluence of the hits (expected distributions studied in \cite{Saldin}), a result of fluctuations in pulse fluence and the positions of particles when illuminated. \label{fig:fluence}}
\end{figure}


We reconstructed 3D Fourier intensities from random starts using EMC with only diffraction data while imposing Friedel symmetry since these are the minimal constraints expected in future cryptotomography experiments. We later evaluated each converged reconstruction (Figs. \ref{fig:recon}) by fitting them to intensity distributions of ideal ellipsoidal particles $I_{\text{ellip}}$. 

We began each reconstruction with random intensities $W({\bf q})$, represented by random numbers on a cubic array. Intensities were reconstructed only in the range $\qmin <   |{\bf q}|  \leq \qmax$ (non-redacted regions in 2D sections of Fig. \ref{fig:recon}) with $\qmin/\qmax=20/44$ --- determined from low spatial frequencies missing-signal region and maximum scattering angle of $5.23^\circ$ in processed hits (white circle in Fig. \ref{fig:data}, lower left).

\begin{figure}[t!]
\centering
\includegraphics[width=2.15in]{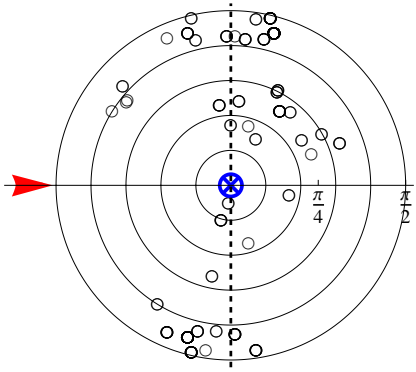}
\caption{(Color online) We superimpose the most likely orientations of the nano-particle symmetry axis in the 54 hits found in each of 10 reconstructions in this azimuthal projection (marked by equidistant rings of constant latitude in this top hemisphere). Orientations are inversion-symmetric to those in the complementary bottom hemisphere. Many orientations coincide. The vertical dashed line shows the detector plane; FEL pulses traveled with the red arrow; nano-particles were injected into the page (blue crosshair).
\label{fig:rotBias}}
\end{figure}

We normalized each random initial $W({\bf q})$ to have tomograms that matched the mean power received per hit; initial fluences were set to $ \sum_i K_{ik}$; we used a $\Mrot=3240$ tomographic representation of our model ($\Mrot$ sufficiency discussed in \cite{Loh}). The noise parameter $\sigma$ was measured in units of $\Delta_K$: the square root of the sum of variances in measured intensities of each pixel \footnote{Undesirable rotational averaging occurs at $\sigma = \Delta_K$. Such reconstructions resemble 3D powder diffraction pattern.}. We determined $\sigma_{\text{min}} = 0.07 \, \Delta_K$, where reconstructions below this showed diminished likelihood and significant diversity \footnote{Reconstructions using $\sigma_{\text{min}}$ routinely converged to one of two quantitatively distinguishable varieties: log-likelihood of one was about $1\%$ higher than the other. We rejected the lower-likelihood variety as candidate solutions. Such multiplicity is expected since reconstructions near $\sigma_{\text{min}}$ were only marginally constrained by so few hits (Fig. \ref{fig:rotBias}).}.

We could in principle recover the nano-particle's real-space contrast from the reconstructed 3D intensities $W$ via phase retrieval  \cite{GoldPyramid}, but there will be pixellation effects (38.3 nm half-resolution in processed hits). Instead we determined the principal radii of the nano-particle from our reconstructed intensities using R-factor comparisons \footnote{R-factor between $I_{\text{ellip}}({\bf q})$ and reconstructed $W$ were computed as $\left( \sum_{\bf q }   \left|I_{\text{ellip}}({\bf q})^{1/2} - W({\bf q})^{1/2} \right|  \right) / \left( \sum_{\bf q } W({\bf q})^{1/2} \right)$, where $I_{\text{ellip}}({\bf q}) \propto \left|  \left( \sin (\pi |{\bf \tilde{q}}|) - \pi |{\bf \tilde{q}}| \cos (\pi |{\bf \tilde{q}}|)  \right)\, /\,  |{\bf \tilde{q}}|^3 \right|^2$; $|{\bf \tilde{q}}| = \sqrt{q_x^2 x_0^2 + q_y^2 y_0^2 + q_z^2 z_0^2 }\, /\qmax  $;  $\qmin <   |{\bf q}|  \leq \qmax$; $x_0, y_0, z_0$ are the principal radii of the nano-particle. $I_{\text{ellip}}({\bf q})$ and $W$ were normalized to the same total power for R-factor comparisons. \label{foot:Rfactor}} with those of an ideal ellipsoidal particle $I_{\text{ellip}}({\bf q})$. The particle's principal radii were found to be $ 30.5 \pm 0.8: 30.6 \pm 0.7 : 76.1 \pm 1.1 $ nm with R-factor $0.093 \pm 0.006$ ($0.162 \pm 0.003$ when reconstructions were compared to the intensity function expected of the SEM particle measurements) \footnote{We reconstructed ten intensity distributions (e.g. Fig. \ref{fig:recon}) from random initial intensities. For each reconstruction we found the principal radii that gave the best R-factor fit to an ellipsoidal intensity $I_{\text{ellip}}({\bf q})$; the quoted radii are their averages.}. We expect a larger error in the longer direction of the prolate nano-particle since it corresponds to the compact direction of its oblate intensities, which was more susceptible to background noise. Fig. \ref{fig:fluence} shows the concomitant fluence distribution we reconstructed.

Fig. \ref{fig:rotBias} shows the most likely orientations of the particles corresponding to the 54 hits used in intensity reconstruction \footnote{EMC found the particle's axial symmetry without imposition. For each hit, EMC automatically assigned nearly equal probability to orientations related by this symmetry, hence mitigating data-scarcity effects expected from limited hits, pixel excision and orientation biases.}. Reconstructed orientational bias in the data could arise from either systematic effects in particle delivery or biases during hit-selection. If a nano-particle's axis of symmetry is colinear with the incident direction of FEL pulses its diffraction pattern will not have identifiable lobes (unlike Fig. \ref{fig:data}). Alternatively, data may contain intensity lobes but are obscured by the redacted pixels or background noise. Such data might have been missed during hit-selection.

Despite the simplicity of our nano-particle, we emphasize that reconstructing its 3D intensities, using only diffraction data, is non-trivial primarily because of ambiguities from unmeasured data orientation and fluence. These ambiguities make direct interpretation of the data very challenging. EMC circumvents these ambiguities without prior assumptions about the intensity distribution beyond Friedel symmetry, enforcing only simple statistics, Eq. \eqref{eqn:Gaussian}, to determine that the particle was prolate instead of oblate. Subsequent R-factor fits of converged EMC reconstructions also gave reasonable particle dimensions,  corroborating the effectiveness of EMC on experimental cryptotomography data. 


Our work was supported by: Helmholtz Association; the PULSE Institute at the SLAC National Accelerator Laboratory by the U.S. Department of Energy, Office of Basic Energy Sciences (MJB); Lawrence Livermore National Laboratory (LLNL) under contracts W-7405-Eng-48 and DE-AC52-07NA27344 ; Laboratory of Directed Research and Development Program of LLNL pertaining to project 05-SI-003; Deutsches Elektronen-Synchrotron, DFG Cluster of Excellence at Munich Centre for Advanced Photonics, Virtual Institute Program of Helmholtz Society, Joachim Herz Stiftung, Max Planck Society and Swedish Research Council. We thank the reviewers for their insightful suggestions and N. D. Loh thanks Yoav Kallus for his invaluable discussions.

\bibliographystyle{unsrt}
\bibliography{nanorice}

\begin{thebibliography}{10}

\bibitem{Bogan2010}
M.~J. Bogan et~al.
\newblock {\em Aerosol Sci. and Tech.}, 44(3):i--iv, 2010.

\bibitem{Neutze}
R.~Neutze et~al.
\newblock {\em Nature}, 406:752--757, 2000.

\bibitem{Chapman2006}
H.~N. Chapman et~al.
\newblock {\em Nature Physics}, 2:839--843, 2006.

\bibitem{Elser}
V.~Elser.
\newblock {\em IEEE Trans. Info. Theory}, 55:4715--4722, 2009.

\bibitem{Loh}
N.~D. Loh and V.~Elser.
\newblock {\em Phys. Rev. E}, 80:026705, 2009.

\bibitem{Ourmazd}
R.~Fung et~al.
\newblock {\em Nature Physics}, 5:64--67, 2009.

\bibitem{GoldPyramid}
H.~N. Chapman et~al.
\newblock {\em J. Opt. Soc. Am. A}, 23(5), May 2006.

\bibitem{beamSize}
J.~Chalupsky et~al.
\newblock {\em Optics Express}, 15(10), 2007.

\bibitem{Bogan2009}
M.~J. Bogan et~al.
\newblock unpublished.

\bibitem{multilayerMirror}
S.~Bajt et~al.
\newblock {\em Appl. Optics}, 47:1673--1683, 2008.

\bibitem{lensStack}
W.~H. Benner et~al.
\newblock {\em J. Aerosol Sci.}, 39:917--928, 2008.

\bibitem{Saldin}
E.~L. Saldin et~al.
\newblock Springer-Verlag, Berlin, 1999.

\bibitem{EM}
A.~P. Dempster et~al.
\newblock {\em J. Royal Stat. Soc., B}, 39:1--38, 1977.

\end{thebibliography}

\end{document}